\documentclass[preprint,aps,nofootinbib]{revtex4}
\usepackage[dvips]{graphicx}
\usepackage{amssymb}
\usepackage{amsmath}
\usepackage{amsmath, amsthm, amssymb, mathrsfs}
\usepackage{pstricks}

\newcommand{\ket}[1]{\lvert #1 \rangle}
\newcommand{\bra}[1]{\langle #1 \lvert}

\begin{document}

\title{Effect of Decoherence in Ekert-Protocol}
\author{
DaeKil Park$^{1}$, Jin-Woo Son$^{2}$, Seong-Keuck Cha$^{3}$, Eun Deock Seo$^{4}$} 

\vspace{1.0cm}

\affiliation{
$^1$ Department of Physics, Kyungnam University, Masan, 631-701, 
Korea  \\
$^2$ Department of Mathematics, Kyungnam University, Masan,
631-701, Korea         \\
$^3$ Department of Chemistry, Kyungnam University, Masan, 631-701, Korea  \\
$^4$ Department of Fire Prevention and Disaster Engineering, Kyungnam University, Masan, 
631-701, Korea}

\begin{abstract}
We have examined the effect of the decoherence in the Ekert91 quantum cryptographic protocol.
In order to explore this issue we have introduced two major decoherences, the depolarizing
channel and the generalized amplitude damping, between the singlet source and one of the 
legitimate users. It is shown that the depolarizing channel disentangles the quantum
channel more easily than the generalized amplitude damping. This fact indicates that the 
Ekert protocol is more robust to the generalized amplitude damping. We also have computed the 
Bell inequality to check the robustness or weakness of the Ekert91 protocol. 
Computation of the Bell
inequality also confirms the robustness of the Ekert91 protocol to the generalized amplitude
damping compared to the depolarizing channel.

\end{abstract}


\maketitle

The purpose of the quantum key distribution (QKD) is to establish a secret key between the
two legitimate users, usually called Alice and Bob, who are assumed to be distant from 
each other. The most well-known protocol for QKD process is BB-84\cite{bb84} scheme. Since
Alice and Bob share the secret key in this scheme using two conjugate bases 
$\{\ket{x}, \ket{y}\}$ and
$\{\ket{u}, \ket{v}\}$, where
\begin{equation}
\label{conjugate1}
\ket{u} = \frac{1}{\sqrt{2}} \left( \ket{x} + \ket{y} \right)    \hspace{2.0cm}
\ket{v} = \frac{1}{\sqrt{2}} \left( \ket{x} - \ket{y} \right),
\end{equation}
BB84 is conventionally called two-bases protocol.

Since the most important issue in the QKD is a security problem, it is important to understand 
the possible strategies of the eavesdropping. The various eavesdropping strategies, therefore, 
were investigated\cite{hutt94,lutk96,gisin97,fuchs97} in the BB84 scenario. Especially, in 
Ref.\cite{fuchs97}, the optimal strategy for the translucent eavesdropping was analytically
analyzed. Following Ref.\cite{fuchs97}, we know that 
the optimal mutual information between sender (Alice)
and eavesdropper (Eve) becomes 
\begin{equation}
\label{optimal1}
{\cal I}_{xy} = \frac{1}{2} \phi \left[2 \sqrt{D_{uv} (1 - D_{uv})} \right]  \hspace{1.0cm}
{\cal I}_{uv} = \frac{1}{2} \phi \left[2 \sqrt{D_{xy} (1 - D_{xy})} \right],
\end{equation}
where the subscripts denote the basis Alice sends a signal to Bob, and 
$\phi(z) = (1+z) \log_2 (1+z) + (1-z) \log_2 (1-z)$. The constants $D_{xy}$ and $D_{uv}$ 
denote the disturbance in these bases. Although the 
optimal eavesdropping strategy can be, in general,
performed with Eve's two qubit probe\cite{griffi97}, authors in Ref.\cite{biham97} has shown 
that it is also possible with single-qubit probe. Recently, the situation that many
eavesdroppers attack the BB84 protocol is discussed\cite{eylee09-1}.

Besides BB84 protocol many different protocols have been suggested to establish the secure
QKD. In Ref.\cite{ekert91} Ekert suggested a different protocol based on the maximally 
entangled EPR state. In this protocol, called Ekert91 protocol, the presence of the 
eavesdroppers can be realized by the legitimate users if they compute the Bell
inequality\cite{bell64,chsh69}. Few years later the implementation and the possible
eavesdropping strategies for the Ekert91 protocol were experimentally realized 
in Ref.\cite{naik99}. Another protocol suggested in Ref.\cite{bruss98} 
(see also Ref.\cite{bech99}) is that Alice sends a signal to Bob using three bases
$\{ \ket{x}, \ket{y} \}$, $\{\ket{u}, \ket{v}\}$ and $\{\ket{w}, \ket{z}\}$, where
\begin{equation}
\label{conjugate2}
\ket{w} = \frac{1}{\sqrt{2}} \left(\ket{x} + i \ket{y} \right)    \hspace{2.0cm}
\ket{z} = \frac{1}{\sqrt{2}} \left(\ket{x} - i \ket{y} \right).
\end{equation}
In this protocol the maximal mutual information between Alice and Eve reduces to 
\begin{equation}
\label{optimal2}
{\cal I}^{AE} = 1 + (1 - D) \bigg\{ f(D) \log f(D) + \left[1 - f(D)\right] 
\log \left[1 - f(D)\right] \bigg\}
\end{equation}
where $D$ is an error rate and 
\begin{equation}
\label{optimal3}
f(D) = \frac{1}{2} \left[ 1 + \frac{1}{1 - D} \sqrt{D (2 - 3 D)} \right].
\end{equation}
Since ${\cal I}^{AE}$ in Eq.(\ref{optimal2}) is less than the two-bases optimal mutual 
information given in Eq.(\ref{optimal1}), we can say that the potocol with three-bases is 
more secure than the usual two-bases BB84 protocol. In addition, the quantum cryptography with
qudit states\cite{bech00,bruss02,boure01,cerf01}, continuum states\cite{piran08}, and 
noisy channels\cite{shad08} are currently investigated.

\begin{figure}[ht!]
\begin{center}
\includegraphics[height=5.0cm]{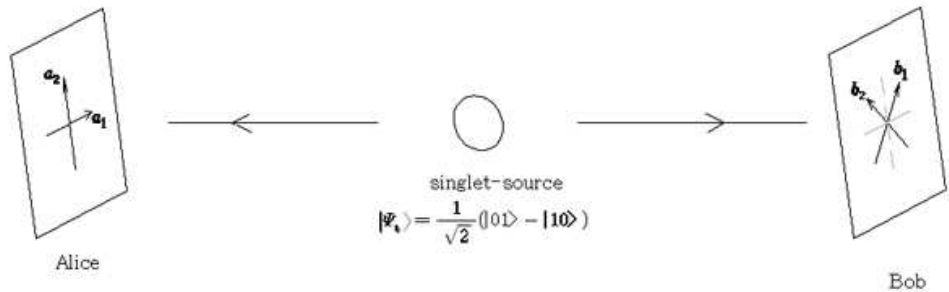}
\caption[fig1]{The schematic diagram for the Ekert91 protocol. The singlet source 
distributes the first qubit to Alice and the second one to Bob. In order to check the 
presence or absence of the eavesdroppers Alice measures the correlation coefficients 
along $\mathbf{a}_1$ and $\mathbf{a}_2$, and Bob along $\mathbf{b}_1$ and $\mathbf{b}_2$.
These measurements enable them to compute the Bell-inequality.}
\end{center}
\end{figure}

In this paper we would like to examine the effect of the decoherence in the Ekert 
protocol\cite{ekert91}. In this protocol the source of the singlet state 
$\ket{\psi_S} = (1 / \sqrt{2}) (\ket{01} - \ket{10})$ sends the first qubit to Alice and 
the second one to Bob. Then, they can establish the QKD via the usual von Neumann measurement.
The presence of the eavesdropper can be readily realized in this protocol by computing the 
Bell inequality. We will examine the effect of the decoherence by introducing the various 
noises between the singlet-source and Bob.

We first consider the depolarizing channel whose quantum operation\cite{nielsen00} is 
\begin{equation}
\label{depol-1}
\varepsilon (\rho) = p \frac{I}{2} + (1 - p) \rho.
\end{equation}
The operator-sum representation of the depolarizing channel can be written as 
\begin{equation}
\label{depol-2}
\varepsilon (\rho) = \sum_{k=0}^3 E_k \rho E_k^{\dagger}
\end{equation}
where the operation elements are given by 
\begin{equation}
\label{depol-3}
E_0 = \sqrt{\frac{1 - 3 p}{4}} I \hspace{1.0cm}
E_1 = \frac{\sqrt{p}}{2} X       \hspace{1.0cm}
E_2 = \frac{\sqrt{p}}{2} Y       \hspace{1.0cm}
E_3 = \frac{\sqrt{p}}{2} Z. 
\end{equation}
In Eq.(\ref{depol-3}) $X$, $Y$, and $Z$ denote the usual Pauli matrices. Since we assume that 
the depolarizing channel is introduced between the singlet-source and Bob, the effect of this
noise changes the singlet state into 
\begin{equation}
\label{depol-4}
\varepsilon_{DP} (\rho_S) = \sum_{k=0}^3 (\openone \otimes E_k) \rho_S 
(\openone \otimes E_k)^{\dagger}  
\end{equation}
where the subscript `DP' stands for depolarizing channel and 
\begin{eqnarray}
\label{depol-5}   
\rho_S \equiv \ket{\psi_S} \bra{\psi_S} = \frac{1}{2}
\left(           \begin{array}{cccc}
           0  &  0  &  0  &  0      \\
           0  &  1  &  -1  &  0     \\
           0  &  -1  &  1  &  0     \\
           0  &  0  &  0  &  0
                  \end{array}                  \right).
\end{eqnarray} 
Inserting the operation elements in Eq.(\ref{depol-3}) into Eq.(\ref{depol-4}), one can easily
derive 
\begin{eqnarray}
\label{depol-6}
\varepsilon_{DP} (\rho_S) = \frac{1}{2}
\left(             \begin{array}{cccc}
      p/2  &  0  &  0  &  0           \\
      0    &  1 - p/2  &  -(1-p)  &  0  \\
      0  &  -(1-p)  &  1 - p/2  &  0    \\
      0  &  0  &  0  &  p/2            
                   \end{array}               \right).
\end{eqnarray} 
At this stage let us assume that the legitimate users, Alice and Bob, want to perform the 
quantum mechanical measurement to realize which states they have. In other words, they perform
a von Neumann measurement with a set of measurement operators
$\{ M_{00}, M_{01}, M_{10}, M_{11} \}$, where
\begin{eqnarray}
\label{depol-7}
& & M_{00} = \ket{0}_A \bra{0} \otimes \ket{0}_B \bra{0}   \hspace{1.0cm}
    M_{01} = \ket{0}_A \bra{0} \otimes \ket{1}_B \bra{1}         
                                                                   \\   \nonumber
& & M_{10} = \ket{1}_A \bra{1} \otimes \ket{0}_B \bra{0}   \hspace{1.0cm}
    M_{11} = \ket{1}_A \bra{1} \otimes \ket{1}_B \bra{1}.
\end{eqnarray}
In Eq.(\ref{depol-7}) the subscripts `A' and `B' correspond to Alice and Bob respectively. 
Making use of the quantum mechanical postulate\cite{nielsen00} it is easy to show that the 
probabilities $P(i, j)$ for outcome $i$ and $j$ are
\begin{equation}
\label{depol-8}
P(0,0) = P(1,1) = \frac{p}{4}   \hspace{2.0cm}
P(0,1) = P(1,0) = \frac{1}{2} - \frac{p}{4}.
\end{equation}
Therefore, the depolarizing channel induces an error in the QKD between Alice and Bob, whose
rate is 
\begin{equation}
\label{depol-9}
D = P(0,0) + P(1,1) = \frac{p}{2}.
\end{equation}
In terms of the error rate $D$, $\varepsilon_{DP} (\rho_S)$ can be written as
\begin{eqnarray}
\label{depol-10}
\varepsilon_{DP} (\rho_S) = \frac{1}{2}
\left(              \begin{array}{cccc}
           D  &  0  &  0  &  0                   \\
           0  & 1-D  &  2D - 1  & 0              \\
           0  &  2D-1  &  1-D  &  0              \\
           0  &  0  &  0  &  D
                    \end{array}                      \right).
\end{eqnarray}

Since Ekert protocol relies on the entanglement of the quantum state emitted by the source, it
is important to keep the high entanglement throughout the procedure. 
If, therefore, the noises disentangle the 
quantum channel, Ekert protocol becomes useless for the creation of QKD. In this reason
in order to 
know whether or not the Ekert protocol is robust under the depolarizing channel
it is important to compute the entanglement of $\varepsilon_{DP} (\rho_S)$. In this paper we 
adopt the concurrence as an entanglement measure, whose computational technique was 
developed in Ref.\cite{wootters98}. Following Ref.\cite{wootters98} it is easy to show that the 
concurrence ${\cal C}_{DP}$ of $\varepsilon_{DP} (\rho_S)$ is 
\begin{eqnarray}
\label{depol-11}
{\cal C}_{DP} = \left\{            \begin{array}{cc}
                        1 - 3 D     & \hspace{1.0cm}   0 \leq D \leq 1/3    \\
                          0         & \hspace{1.0cm}   1/3 \leq D \leq 1/2.
                                   \end{array}                  \right.
\end{eqnarray}
If, therefore, $D$ is too large, ${\cal C}_{DP}$ is too small which means that Ekert protocol
becomes useless.

Another factor we should carefully examine is the Bell-inequality. Since Alice and Bob assume
the presence of the eavesdroppers if the Bell-inequality is not violated, it is important to
understand the effect of noises on the Bell-inequality. In order to discuss this issue we 
consider first the correlation coefficient of the measurement performed by Alice along
$\mathbf{a}_i$ and by Bob along $\mathbf{b}_j$:
\begin{equation}
\label{bell-1}
E(\mathbf{a}_i, \mathbf{b}_j) = P_{++} (\mathbf{a}_i, \mathbf{b}_j) 
                                        + P_{--} (\mathbf{a}_i, \mathbf{b}_j)
- P_{+-} (\mathbf{a}_i, \mathbf{b}_j) - P_{-+} (\mathbf{a}_i, \mathbf{b}_j)
\end{equation}
where $P_{\pm \pm} (\mathbf{a}_i, \mathbf{b}_j)$ denotes the probability that result $\pm$ has
been obtained along $\mathbf{a}_i$ and $\pm$ along $\mathbf{b}_j$. 
Then, the CHSH expression of the 
Bell-inequality is $|S| \leq 2$, where 
\begin{equation}
\label{bell-2}
S = E(\mathbf{a}_1, \mathbf{b}_1) - E(\mathbf{a}_1, \mathbf{b}_2) + 
E(\mathbf{a}_2, \mathbf{b}_1) + 
E(\mathbf{a}_2, \mathbf{b}_2).
\end{equation}
In Ekert protocol the measurement directions $\mathbf{a}_i$ and $\mathbf{b}_j$ lie in
the $x$-$y$ plane, perpendicular to the trajectory of the particles, with azimuthal angles
$\theta_i$ and $\theta_j$. Then, $P_{\pm \pm} (\mathbf{a}_i, \mathbf{b}_j)$ for the quantum 
state $\rho$ can be computed as follows:
\begin{eqnarray}
\label{bell-3}
& & P_{++} (\mathbf{a}_i, \mathbf{b}_j) = \mbox{Tr} \left[ M_{++}^{\dagger} M_{++} \rho \right]
                                                                          \hspace{1.0cm}
P_{+-} (\mathbf{a}_i, \mathbf{b}_j) = \mbox{Tr} \left[ M_{+-}^{\dagger} M_{+-} \rho \right]
                                                                     \\    \nonumber
& & P_{-+} (\mathbf{a}_i, \mathbf{b}_j) = \mbox{Tr} \left[ M_{-+}^{\dagger} M_{-+} \rho \right]
                                                                          \hspace{1.0cm}
P_{--} (\mathbf{a}_i, \mathbf{b}_j) = \mbox{Tr} \left[ M_{--}^{\dagger} M_{--} \rho \right]
\end{eqnarray}
where the measurement operators are
\begin{eqnarray}
\label{bell-4}
& &M_{++} = \ket{+,\mathbf{a}_i}_A \bra{+,\mathbf{a}_i} \otimes \ket{+,\mathbf{b}_j}_B 
                        \bra{+,\mathbf{b}_j}       \hspace{.5cm}
M_{+-} = \ket{+,\mathbf{a}_i}_A \bra{+,\mathbf{a}_i} \otimes \ket{-,\mathbf{b}_j}_B 
                        \bra{-,\mathbf{b}_j}
                                                                      \\   \nonumber
& &M_{-+} = \ket{-,\mathbf{a}_i}_A \bra{-,\mathbf{a}_i} \otimes \ket{+,\mathbf{b}_j}_B 
                        \bra{+,\mathbf{b}_j}       \hspace{.5cm}
M_{--} = \ket{-,\mathbf{a}_i}_A \bra{-,\mathbf{a}_i} \otimes \ket{-,\mathbf{b}_j}_B 
                        \bra{-,\mathbf{b}_j}.
\end{eqnarray}
In Eq.(\ref{bell-4}) 
$\ket{\pm, \mathbf{a}_i} = (1/\sqrt{2}) (\ket{0} \pm e^{i \theta_i} \ket{1})$  and 
$\ket{\pm, \mathbf{b}_j} = (1/\sqrt{2}) (\ket{0} \pm e^{i \theta_j} \ket{1})$. If $\rho$ is 
noiseless singlet state, it is easy to show that Eq.(\ref{bell-3}) gives 
\begin{eqnarray}
\label{bell-5}
& & P_{++} (\mathbf{a}_i, \mathbf{b}_j) = P_{--} (\mathbf{a}_i, \mathbf{b}_j) 
= \frac{1}{4} \left[1 - \cos (\theta_i - \theta_j) \right]           \\  \nonumber 
& & P_{+-} (\mathbf{a}_i, \mathbf{b}_j) = P_{-+} (\mathbf{a}_i, \mathbf{b}_j)   
= \frac{1}{4} \left[1 + \cos (\theta_i - \theta_j) \right] 
\end{eqnarray}
and, as a result, $E(\mathbf{a}_i, \mathbf{b}_j) = - \mathbf{a}_i \cdot \mathbf{b}_j$.
If we choose the azimuthal angles of $\mathbf{a}_1$, $\mathbf{a}_2$, $\mathbf{b}_1$ and 
$\mathbf{b}_2$ as $0$, $\pi/2$, $\pi/4$ and $3\pi/4$, we have $|S| = 2 \sqrt{2}$. Thus, 
the Bell-inequality is violated for the noiseless case, which can be used to check the 
presence of eavesdroppers in Ekert protocol.

For the case of the depolarizing channel the quantum state $\rho$ should be changed into
$\varepsilon_{DP} (\rho_S)$ in Eq.(\ref{depol-10}). In this case it is easy to show that
the factor $S$ becomes
\begin{equation}
\label{depol-12}
|S_{DP}| = 2 \sqrt{2} (1 - 2 D).
\end{equation}
Therefore, the Bell-inequality is violated in the region $0 \leq D \leq D_c^{DP}$, where
the critical error rate $D_c^{DP}$ is 
\begin{equation}
\label{depol-13}
D_c^{DP} = \frac{1}{2} \left[1 - \frac{\sqrt{2}}{2} \right] \sim 0.146.
\end{equation}
When $D > D_c^{DP}$, the Bell-inequality is satisfied and the Ekert protocol via the 
depolarizing channel becomes useless. Combining Eq.(\ref{depol-11}) and Eq.(\ref{depol-12}), 
one can say that the Ekert protocol is not useful for the creation of QKD when the 
concurrence is less than ${\cal C}_{DP}^c$, where the critical concurrence is 
\begin{equation}
\label{depol-14}
{\cal C}_{DP}^c = 1 - 3 D_c^{DP} = \frac{3 \sqrt{2} - 2}{4} \sim 0.51.
\end{equation}

It is worthwhile noting that the choice of the measurement direction on the $x$-$y$ plane 
for the noiseless case
does give the optimal $S$-factor for the case of the depolarizing channel too. This can be 
easily shown following Ref.\cite{horo95-1}. However, this does not hold in general. 
For example, let us consider the bit-flip, whose quantum operation is 
\begin{equation}
\label{bf-1}
\varepsilon_{BF} (\rho_S) = \sum_{k=0}^1 (\openone \otimes E_k) \rho_S
(\openone \otimes E_k)^{\dagger}
\end{equation}
with
\begin{equation}
\label{bf-2}
E_0 = \sqrt{p}  I     \hspace{2.0cm}   E_1 = \sqrt{1-p} X.
\end{equation}
In this case $\varepsilon_{BF} (\rho_S)$ becomes finally 
\begin{eqnarray}
\label{bf-3}
\varepsilon_{BF} (\rho_S) = \frac{1}{2} 
\left(            \begin{array}{cccc}
          D  &  0  &  0  &  -D                 \\
          0  &  1-D  &  -(1 - D)  &  0         \\
          0  &  -(1-D)  &  1 - D  &  0         \\
          -D  &  0  &  0  &  D
                   \end{array}                          \right)
\end{eqnarray}
where $D = 1 - p$. Then it is easy to show that the above-mentioned calculation procedure 
gives $S$-factor $|S_{BF}| = 2 \sqrt{2} (1 - D)$ while the optimal $S$-factor is 
$|S| = 2 \sqrt{2} \sqrt{(1 - D)^2 + D^2}$. In this case, therefore, the choice of the 
measurement direction on $x$-$y$ plane for the noiseless case 
does not give the optimal $S$-factor in the presence of the decoherence..

Finally, we would like to discuss the Ekert protocol with the generalized amplitude 
damping, whose operation elements are 
\begin{eqnarray}
\label{gad-1}
& & E_0 = \sqrt{p} \left(        \begin{array}{cc}
                            1  &  0                    \\
                            0  &  \sqrt{1 - \gamma}    
                                 \end{array}                   \right)         \hspace{1.0cm}
   E_1 = \sqrt{p} \left(         \begin{array}{cc}
                            0  &  \sqrt{\gamma}        \\
                            0  &  0
                                  \end{array}                 \right)  
                                                                            \\   \nonumber
& & E_2 = \sqrt{1-p} \left(        \begin{array}{cc}
                            \sqrt{1 - \gamma}  &  0                    \\
                            0  &  1    
                                 \end{array}                   \right)         \hspace{1.0cm}
   E_3 = \sqrt{1-p} \left(         \begin{array}{cc}
                            0  &  0        \\
                            \sqrt{\gamma}  &  0
                                  \end{array}                 \right).
\end{eqnarray}
The amplitude damping is one of the important quantum noise, which describes the effect of 
energy dissipation. Making use of Eq.(\ref{gad-1}), one can show straightforwardly that the 
quantum operation $\varepsilon_{GAD} (\rho_S)$ reduces to
\begin{eqnarray}
\label{gad-2}
\varepsilon_{GAD} (\rho_S) = \frac{1}{2}     
        \left(                  \begin{array}{cccc}
             2 p D  &  0  &  0  &  0                                    \\
             0  &  1 - 2pD  &  -\sqrt{1 - 2 D}  &  0                    \\
             0  &  -\sqrt{1 - 2D}  &  1 - 2(1-p) D  &  0                \\
             0  &  0  &  0  &  2 (1-p) D           
                                 \end{array}            \right)
\end{eqnarray}
where the error rate is defined as $D = \gamma / 2$. Following Wootters' 
procedure\cite{wootters98}, one can show that the concurrence ${\cal C}_{GAD}$ for the mixed 
state (\ref{gad-2}) becomes
\begin{eqnarray}
\label{gad-3}
{\cal C}_{GAD} = \left\{           \begin{array}{ll}
     \lambda_+ - \lambda_- - \lambda_1 - \lambda_2 & \hspace{1.0cm} D \leq \mu(p)   \\
                         0  &  \hspace{1.0cm}       D \geq \mu(p)       
                                   \end{array}                  \right.
\end{eqnarray}
where
\begin{eqnarray}
\label{gad-4}
& & \mu(p) = \frac{\sqrt{1 + 4p (1 - p)} - 1}{4 p (1 - p)}              \\   \nonumber
& & \lambda_{\pm} = \left(\frac{ [1 - 2 D + 2p(1-p) D^2] \pm \sqrt{(1-2D) 
                                                        [1 - 2D + 4 p (1-p) D^2]} }
                               {2}   \right)^{1/2}                     \\     \nonumber
& & \lambda_1 = \lambda_2 = \sqrt{p (1-p)} D.
\end{eqnarray}
It is worthwhile noting that ${\cal C}_{GAD}$ has symmetry under the $p \leftrightarrow (1-p)$
interchange.

\begin{figure}[ht!]
\begin{center}
\includegraphics[height=8.0cm]{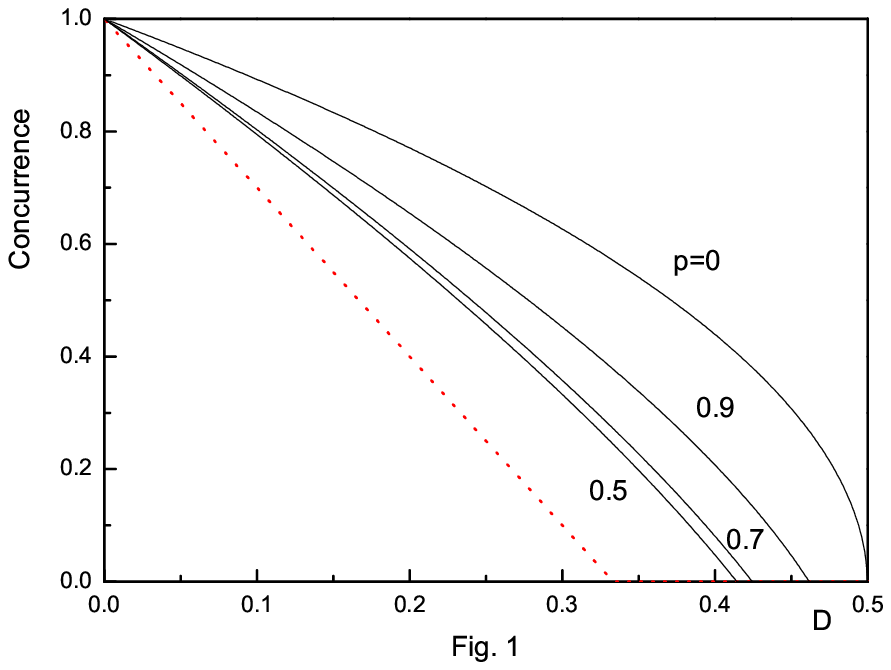}
\caption[fig2]{The plot of $D$-dependene of the concurrences. The dotted line is the concurrence
for the depolarizing channel given in Eq.(\ref{depol-11}). The solid lines are concurrences
for the generalized amplitude damping given in Eq.(\ref{gad-3}) with $p=0$, $0.9$, $0.7$ and 
$0.5$. Since the dotted line is less
than other concurrences, this fact implies that the Ekert protocol is weak under the 
depolarizing channel compared to the generalized amplitude damping.}
\end{center}
\end{figure}

Fig. 1 shows the $D$-dependence of the concurrence ${\cal C}_{GAD}$ 
with $p=0$, $0.9$, $0.7$ and $0.5$. For comparison we plot the concurrence for the 
depolarizing channel together. Since the concurrence for the depolarizing channel is less than
those for the generalized amplitude damping, this means that the quantum channel for the 
Ekert protocol can be disentangled more easily in the depolarizing channel. Therefore, we 
can conclude that the Ekert protocol is more robust to the generalized amplitude damping
than the depolarizing channel. Among the generalized amplitude damping the Ekert 
protocol is weak at $p=0.5$ and robust at $p=0$ or $1$.

The $S$-factor for the generalized amplitude damping can be straightforwardly computed.
It is interesting that the $S$-factor is independent of $p$ as follow:
\begin{equation}
\label{gad-5}
| S_{GAD} | = 2 \sqrt{2} \sqrt{1 - D}.
\end{equation}
This $S$-factor is optimal even in the presence of the decoherence like the depolarizing
channel\cite{horo95-1}. The Bell-inequality is violated in the region $0 \leq D \leq D_c^{GAD}$,
where the critical error rate $D_c^{GAD}$ is 
\begin{equation}
\label{gad-6}
D_c^{GAD} = \frac{1}{4} = 0.25.
\end{equation}
Since $D_c^{GAD} > D_c^{DP}$, it also indicates that the Ekert protocol is more robust for the 
generalized amplitude damping than for the depolarizing channel.

In this paper the effects of the depolarizing channel and the generalized amplitude damping
are investigated in the Ekert91 quantum cryptographic protocol. The decrease of the 
entanglement of the quantum channel has been explicitly calculated when the decoherences are
introduced between the singlet state source and Bob. The decreasing rate in the depolarizing
channel is much larger than that in the generalized amplitude damping. This fact implies that 
the Ekert91 protocol is more robust to the generalized amplitude damping compared to the
depolarizing channel. 

The Bell-inequality is also computed when the decoherences change the quantum channel into
the mixed state. It is shown that the measurement direction chosen for the optimal 
$S$-factor in the noiseless case is also optimal when the depolarizing channel and the 
generalized amplitude damping are introduced. However, this maintenance does not hold
in general. We have explicitly shown that this maintenance does not hold in the bit-flip.

Combining the computation of the entanglement of the quantum channel and the Bell inequality,
we have shown that the Ekert91 protocol becomes useless when $D \geq D_c^{DP} = 0.146$ for
the depolarizing channel and $D \geq D_c^{GAD} = 0.25$ for the generalized amplitude 
damping, where $D$ is an error rate between Alice and Bob. The fact $D_c^{DP} < D_c^{GAD}$
confirms again that the Ekert91 protocol is more robust to the generalized amplitude damping.

{\bf Acknowledgement}: 
This work was supported by the Kyungnam University
Foundation Grant, 2008.

\end{document}